\documentclass{iau_FM}
\usepackage{graphicx}

\title[FM 8.~~Galaxy outskirts magnetic field] 
{Magnetic field constraint in the outskirts of spiral galaxies}

\author[E. Lopez, J. Armijos, M. Llerena, F. Aldaz]   
{Ericson Lopez$^{1,2}$
 Jairo Armijos$^1$
 Mario Llerena$^1$
 Franklin Aldaz$^1$
 }

\affiliation{$^1$Quito Astronomical Observatory, National Polytechnic School,\\ Box 17-01-165 Quito,  Ecuador\\ email: {\tt ericsson.lopez@epn.edu.ec} \\[\affilskip]
$^2$Physics Department, Sciences Faculty, National Polytechnic School, \\ Box 17-01-2759 Quito, Ecuador}

\pubyear{2018}
\jname{Astronomy in Focus} 
\editors{Teresa Lago, ed.}
\begin{document}

\maketitle

\begin{abstract}
Based on CO(2-1) public data, we study the monoxide oxygen gas excitation conditions and the magnetic field strength of NGC 2841, NGC 3077, NGC 3184, NGC 3351 spiral galaxies. For their galaxy outskirts, we found kinetic temperatures in the range of $35 - 38 ~K$, CO column densities $10^{15} - 10^{16} ~cm^{-2}$ and H2 masses of $4 ~10^6 - 6 ~ 10^8 M_\odot$. An H2 density 1$0^3 cm^{-3}$ is suitable to explain the 2 sigma upper limits of the CO(2-1) line intensity. We constrain the magnetic field strength for our sample of spiral galaxies and their outskirts, evaluating a simplified expression of the magneto-hydrodynamic force equation. Our estimations provide values for the magnetic field strength in the order of $6-31 ~\mu G$.  

\end{abstract}

\firstsection 
\section{Introduction}
In this work, we study the magnetic field strength in the outskirts of four spiral galaxies: NGC 2841, NGC 3077, NGC 3184, NGC 3351, following a different approach to those commonly based on Faraday rotation, dust polarization, synchrotron emission, and so on. To constrain the magnetic field strength of spiral galaxies we will follow the Dotson method (\cite[Dotson 1996]{Dot96}), i.e., approaching the magneto-hydrodynamic force equation to derive a simple expression that let us to estimate the upper limit of the magnetic field.
On the other hand, to estimate the density $n$ and mass $M$ of the sources, we use the carbon monoxide emission as a tracer for the molecular gas H$_2$ (\cite[Neininger et al. 1998]{Nei98}). This is because H$_2$ is invisible in the cold interstellar medium, at temperatures around $10-20 ~K$, so its distribution and motion must be inferred from observations of minor constituents of the clouds, such as carbon monoxide and dust. The carbon monoxide is considered a good tracer for the molecular hydrogen (\cite[Neininger et al. 1998]{Nei98}).

\section{Treatment of Carbon Monoxide Data}
\begin{figure}[h]
\begin{center}
\includegraphics[width=0.45 \linewidth]{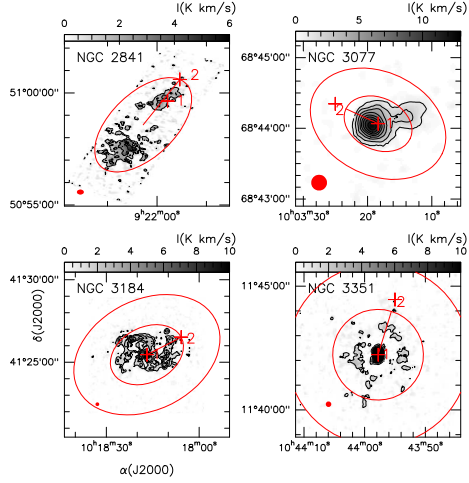}
\caption{CO(2-1) integrated intensity maps for our galaxy sample. The red crosses show the positions used to extract the spectra. The red ellipses indicate the regions used to measure CO(2-1) luminosity. The tiny filled red ellipse represents the IRAM telescope beam (13 arcsec at the CO(2-1), frequency transition of 230.538 GHz). }
\label{fig1}
\end{center}
\end{figure}

To carry out this study we use public CO(2-1) data, first published by  \cite[Leroy et al. (2009)]{Ler09}, data obtained with the IRAM $30 ~m$ telescope located in Spain. 
At the CO(2-1) transition frequency (230.538 GHz), the IRAM telescope has a spatial resolution of 13 arcsec. In Fig. \ref{fig1}, CO(2-1) integrated intensity maps for each galaxy are shown.
To derive $n$ and $M$ from the CO(2-1) data we have selected the spectra by choosing two positions; one located in the nucleus and the second located on the outskirts of the galaxy.

\begin{table}
\begin{footnotesize}
\centering
\begin{tabular}{ccccccc}
\\ \hline
Galaxy name	& Region & $L_{CO}$ & $r$ & $M_{H_2}$ & ($M_{H_2}$+$M_{HI}$) & $B$\\
& &$\times$ 10$^6$ K km s$^{-1}$ pc$^2$ & kpc &$\times$ 10$^7$ M$_{\odot}$	& $\times$ 10$^8$ M$_\odot$	& $\mu$G\\ 
\\ \hline
NGC 2841 &disk         &2.0& 8.5   &103.0&89.6&$\lesssim$31\\
        &outskirts&...& ...  &...  &... &...\\
NGC 3077&disk     &2.3      &0.4   &1.3      &0.4      &$\lesssim$6\\
       &outskirts&$\lesssim$0.7&0.8   &$\lesssim$0.4&$\lesssim$0.1&$\lesssim$7\\
NGC 3184&disk     &188.0    &6.6   &103.0 &36.8&$\lesssim$14\\
        &outskirts&$\lesssim$81.1&13.2&$\lesssim$44.6&$\lesssim$15.9&$\lesssim$19\\
NGC 3351&disk&226.0&7.1 &124.0&32.1&$\lesssim$11\\
        &outskirts&$\lesssim$109.0&14.2&$\lesssim$59.8&$\lesssim$15.5&$\lesssim$15\\
\\ \hline
\end{tabular}
\caption{Parameters derived for the galaxy sample}\label{table3}
\end{footnotesize}
\end{table}

\section{Magnetic fields in the galaxies and their outskirts }

Gaussian fits to the CO(2-1) lines have been performed. Then, the Large Velocity Gradient (LVG) modeling (\cite[van der Tak et al. 2007]{Van07}) is employed to estimate the gas density ($n$), based on the CO(2-1) line width ($\Delta V$) and the line intensity ($I$). To estimate CO integrated luminosity ($L_{CO}$), we use the inner  ellipse for the nuclear region (the $L_{CO}$ will be used later to estimate H$_2$ masses). 
To constrain the magnetic field strength for a given galaxy and its outskirts, we use the expression $B< 3.23\times 10^{-8}\left({R}\over{pc}\right)^{0.5}\left({n}\over{ {cm^{-3}} }\right)^{0.5}\left({M}\over{{M_{\odot}}}\right)^{0.5}\left({r}\over{{pc}}\right)^{-1}$ given by \cite[Dotson (1996)]{Dot96}. This relation includes the $n_{H_2}$, the mass $M$ and radius $r$ of the galaxy, and the radius of curvature $R$ of the magnetic field lines. The $n_{H_2}$ and $M$ were derived using observational data, however the mass that we use here is referred to the dust mass which is obtained by the relation ($M_{H_2}$+$M_{HI}$)/100, i.e. assuming the typical dust-to-gas mass ratio of 0.01.
The molecular hydrogen mass ($M_{H_2}$) for the galaxy disk and its outskirts is derived using the  relation ${{M_{H_2}}\over{M_{\odot}}}=5.5~{{R_{21}}\over{0.8}}\left({L_{CO}}\over{K km  s^{-1}  pc^2}\right)$ given by \cite[Leroy et al. (2009)]{Ler09}, where $R_{21}$ is the CO(2-1)/CO(1-0) line intensity ratio equal to 0.8 and $L_{CO}$ is the CO luminosity. In this relation, a CO(1-0)/H$_2$ conversion factor of $2\times 10^{20}$ cm$^{-1}$ (K km s$^{-1}$)$^{-1}$ has been adopted (\cite[Leroy et al. 2009]{Ler09}).

\section{Discussion and Conclusions}
 In the present contribution, we have estimated the magnetic field strength in the galaxy nuclei and in the outskirts of NGC 2841, NGC 3077, NGC 3184, NGC 3351 spiral galaxies. For that, we have used an approximated expression of the magneto-hydrodynamics to find an upper limit for the magnetic field magnitudes.
 The magnetic field strength lies within the range of $\lesssim$6-31 $\mu$G, which are in good agreement with the values provided by other author for spiral galaxies. A first good idea about the strength of the magnetic field is possible to obtain directly from the estimation of molecular hydrogen mass and radio of the source, without the necessity of a magneto-hydrodynamic model or using a traditional technique like Faraday rotation or Zeeman line broadening. This is a rough estimation that works if the gas pressure is uniform and the viscosity is neglected. A better approach can be obtained keeping more term in the magneto-hydrodynamics force equation (\cite[Dotson 1996]{Dot96}) to impede gravitational collapse.

\end{document}